\documentclass[letter,longauth]{aa}
\usepackage{txfonts,natbib,graphicx}
\usepackage[english]{babel}

\begin{document}
\title{The {\it Herschel} revolution: unveiling the morphology of the
  high mass star formation sites N44 and N63 in the
  LMC\thanks{Herschel is an ESA space observatory with science
    instruments provided by European-led Principal Investigator
    consortia and with important participation from NASA.}}

   \titlerunning{sub-mm view of N44 and N63 in the LMC}

   \author{S.~Hony\inst{1}
     \and F.~Galliano\inst{1}
     \and S.~C.~Madden\inst{1}
     \and P.~Panuzzo\inst{1}
     \and M.~Meixner\inst{2}
     \and C.~Engelbracht\inst{3}
     \and K.~Misselt\inst{3}
     \and M.~Galametz\inst{1}
     \and M.~Sauvage\inst{1}
     \and J.~Roman-Duval\inst{2}
     \and K.~Gordon\inst{2}
     \and B.~Lawton\inst{2}
     \and J.-P.~Bernard\inst{4}
     \and A.~Bolatto\inst{5}
     \and K.~Okumura\inst{1}
     \and C.-H.~R.~Chen\inst{6}
     \and R.~Indebetouw\inst{6}
     \and F.~P.~Israel\inst{7}
     \and E.~Kwon\inst{8}
     \and A.~Li\inst{9}
     \and F.~Kemper\inst{10}
     \and M.~S.~Oey\inst{11}
     \and M.~Rubio\inst{12}
              }

 \institute{
Service d'Astrophysique, CEA, Saclay, 91191 Gif-Sur-Yvette Cedex, France \email{sacha.hony@cea.fr}  \\
\and Space Telescope Science Institute, 3700 San Martin Drive, Baltimore, MD 21218, USA \\
\and Steward Observatory, University of Arizona, 933 North Cherry Ave., Tucson, AZ 85721, USA  \\
\and Centre d'\'{E}tude Spatiale des Rayonnements, CNRS, 9 av. du Colonel Roche, BP 4346, 31028 Toulouse, France  \\
\and Department of Astronomy,  Lab for Millimeter-wave Astronomy, University of Maryland. College Park, MD 20742-2421, USA \\
\and Department of Astronomy, University of Virginia, PO Box 3818, Charlottesville, VA 22903, USA \\
\and Sterrewacht Leiden, Leiden University, PO Box 9513, 2300 RA Leiden, The Netherlands \\
\and Astronomy \& Space Science, Sejong University, 143-747, Seoul, South Korea \\
\and Department of Physics and Astronomy, University of Missouri, 314 Physics Building, Columbia, MO 65211, USA \\
\and Jodrell Bank Centre for Astrophysics, Alan Turing Building, School of Physics and Astronomy, The University of Manchester, Oxford Road, Manchester, M13 9PL, UK \\
\and Department of Astronomy, University of Michigan, 830 Dennison Building, Ann Arbor, MI 48109-1042, USA \\
\and Departamento de Astronomia, Universidad de Chile, Casilla 36-D, Santiago, Chile \\
}

   \date{01/04/2010 ; \today }

  \abstract
  {}
  {We study the structure of the medium surrounding sites of high-mass
    star formation to determine the interrelation between the H~{\sc
      ii} regions and the environment from which they were formed. The
    density distribution of the surrounding{s} is key in determining
    how the radiation of the newly formed stars { interacts with the
      surrounds in a way that allows it to be used as a star formation
      tracer}. }
  {We present new {{\it Herschel/SPIRE}} 250, 350 and 500~$\mu$m data
    of LHA 120-N44 and LHA 120-N63 in the LMC. We construct average
    spectral energy distributions (SEDs) for annuli centered on the IR
    bright part of the star formation sites. The annuli cover
    $\sim$10$-$$\sim$100 pc. We use a phenomenological dust model to
    fit these SEDs to derive the dust column densities, characterise
    the incident radiation field and the abundance of polycyclic
    aromatic hydrocarbon molecules. We see a factor 5 decrease in the
    radiation field {energy} density {as a function of radial
      distance} around N63. N44 does not show a systematic trend. We
    construct a simple geometrical model to derive the 3-D density
    profile of the surroundings of these two regions.}
  {{{\it Herschel/SPIRE}} data {have proven} very efficient in
    deriving the dust mass distribution. We find that the radiation
    field in the two sources behaves very differently. N63 is more or
    less spherically symmetric and the average radiation field drops
    with distance. N44 shows no systematic decrease of the radiation
    intensity which is probably due to the {inhomogeneity} of the
    surrounding molecular material and to the complex distribution of
    several star forming clusters in the region.}
  {}

   \maketitle

\begin{figure*}
  \sidecaption
  \includegraphics[width=6cm]{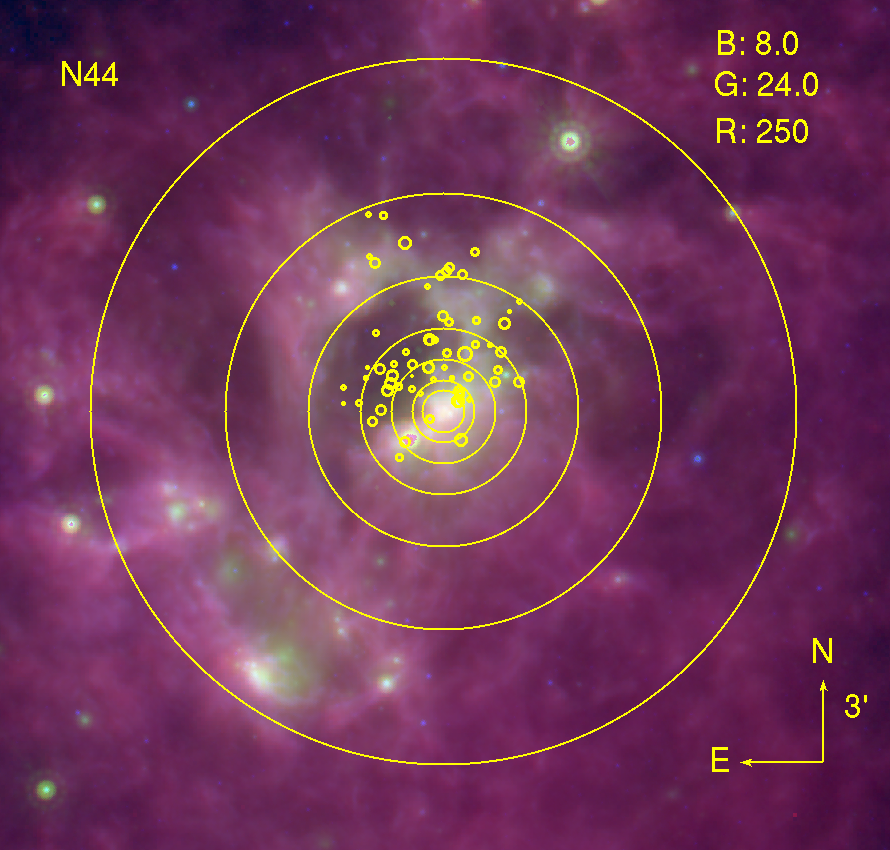}
  \includegraphics[width=6cm]{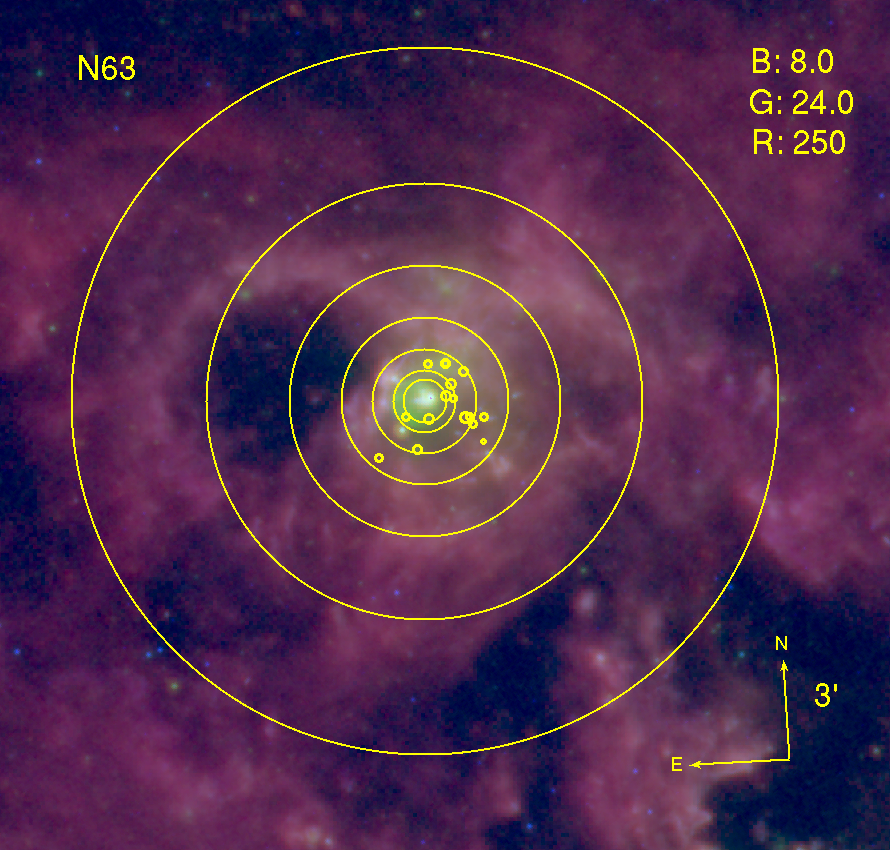}
  \caption{False colour images of the surroundings of N44 and N63.
    IRAC 4 (8.0~$\mu$m; blue) traces the PAH emission, i.e. dominated
    by the PDRs that are the result of the UV radiation impinging on
    the surfaces of the molecular clouds. MIPS 24~$\mu$m (green)
    traces the hottest dust and is generally a good tracer of the
    ionised medium around SF sites. SPIRE 250~$\mu$m (red) maps to
    first order the dust column density and is less sensitive to the
    actual illumination conditions. The circles designate the annuli
    for which we construct SEDs. The coordinates of the centers are
    05:22:03.2 -67:57:55 and 05:35:41.3 -66:01:45 [J2000] for N44 and
    N63, respectively. The yellow circles indicate the positions of
    optically detected ionising stars
    \citep{1995ApJ...452..210O,1996ApJS..104...71O}}
  \label{figMaps}
\end{figure*}

\section{Introduction}
\label{sec:introduction}
High-mass star formation (SF) sites (hereafter HMSFSs) are the beacons
by which we probe a large part of the physics of external galaxies.
They generally represent the most important tracers of the properties
of their host galaxies in terms of star formation rate (SFR) and
general activity. The most frequent tracers of the star-formation
activity generally use the fact that the abundant UV light coming from
the hot, young stars is absorbed in the vicinity and reradiated in the
form of line or continuum emission. This is true, for example, for
H$\alpha$ \citep[e.g.][]{1998ARA&A..36..189K}, the aromatic emission
bands in the mid-IR \citep[e.g.][]{2007ApJ...666..870C} or the IR
continuum due to solid-state materials \citep[dust,
e.g.][]{1986ApJ...303L..41S}. These tracers work relatively well and
are used to characterise nearby star forming regions and star forming
galaxies out to large redshifts, although interesting discrepancies
have been noted for dwarf galaxies at low SFR
\citep[see][]{2009ApJ...706..599L}. One of the main assumptions that
enters into the quantitative interpretation of these data is the
geometry of the material surrounding the newly formed stars, in
particular, where the UV light is being reprocessed. For example, if
the UV photons escape from the ionised medium this may boost the
aromatic feature strengths and strongly influence the lines
originating from the surrounding photo-dissociation regions (PDRs).
There are indications that this geometry in external galaxies may
qualitatively and quantitatively differ from that observed in the
Milky Way (MW)
\citep[e.g.][]{2006A&A...446..877M,2009A&A...508..645G}. One simple
effect may be that at different metallicities the surrounding medium
is more or less opaque and therefore the UV photons have a different
mean free path. More complex scenarios are also discussed in the
literature. For example, clumpiness of the molecular cloud may lead to
small molecular cores surrounded by large PDRs. {\it Herschel} with
its unprecedented wavelength coverage and angular resolution at
submillimeter (submm) wavelengths provides a unique opportunity to
probe the cold interstellar medium (ISM) and sample the effects of the
environment on the resulting SF tracers. In particular, it is well
suited to trace the distribution of matter around HMSFSs and to map
the way the UV radiation permeates and heats the surroundings. Here we
present a study of two HMSFSs in the Large Magellanic Cloud (LMC)
based on data taken in the HERITAGE program \citep[PI. Meixner,
see][]{meixner_special_issue}.

\section{Data treatment and modeling}
\label{sec:data-treatm-model}
We construct spectral energy distributions (SEDs) of the environment
of two distinct HMSFSs (\object{LHA 120-N 44} and \object{LHA 120-N
  63} \citep{1956ApJS....2..315H}, hereafter N44 and N63,
respectively) in the strip of the LMC that was mapped during the
science demonstration phase. The two regions were chosen because they
are the brightest and most isolated regions observed. N44 is the
brightest H~{\sc ii}~complex in the observed strip. Massive star
formation has been active in this region; it contains three OB
associations LH47, 48, and 49 with ages $\gtrsim$ 10 Myr at the
central super-bubble and $\lesssim$ 5 Myr at the super-bubble rims and
surrounding dense H~{\sc ii}~regions, as well as a large number of
massive young stellar objects (YSOs) with ages $\lesssim$ 1 Myr
\citep{1970AJ.....75..171L,1995ApJ...452..210O,2009ApJ...695..511C}.
N44 is also the brightest source at all of the SPIRE wavelengths (250,
350, and 500~$\mu$m) in the strip \citep{meixner_special_issue}. By
contrast, N63 is a simple, roughly round H~{\sc ii}~region. It
contains one OB association LH83 with an age $<$ 5 Myr and a number of
massive YSOs \citep{1996ApJS..102...57B,2008ApJ...678..200C}. At SPIRE
wavelengths, N63 is in a relatively isolated environment. Thus, we use
N44 as an example of a prominent H~{\sc ii}~regions seen in more
distant galaxies and N63 as a comparison whose simple structure makes
it more straightforward to relate dust properties with physical
conditions of the ISM.

The data we use are: {\it 2MASS} J, H, K$_s$
\citep{2006AJ....131.1163S}, {\it Spitzer} IRAC1, 2, 3, 4
\citep{2004ApJS..154...10F}, MIPS 24, 70 and 160 $\mu$m
\citep{2004ApJS..154...25R} and {\it Herschel}
\citep{pilbratt_special_issue} SPIRE 250, 350 and 500 $\mu$m
\citep{griffin_special_issue}. See \citet{meixner_special_issue} for a
description of the SPIRE data treatment. We do not use the PACS
information since the data we have until now does not allow us to
extract extended source fluxes with sufficient accuracy. We have
extracted maps of 40$^{\prime}$$\times$40$^{\prime}$ centered on each
H~{\sc ii} region. We have convolved these data to a spatial
resolution of 38$^{\prime\prime}$ set by the MIPS 160~$\mu$m/SPIRE
500~$\mu$m data. The 2MASS data have been convolved with the beam of
MIPS 160~$\mu$m, the IRAC1$-$4, MIPS 24, 70~$\mu$m have been convolved
using custom-made kernels \citep{2008ApJ...682..336G}. The SPIRE data
have all been convolved to the SPIRE 500~$\mu$m resolution assuming
Gaussian beam profiles with FWHM of 18.1, 25.2 and
36.9$^{\prime\prime}$ for SPIRE 250, 350 and 500, respectively. We
have resampled the convolved images to the pixel scheme of the MIPS
160~$\mu$m image using the IDL/astrolib routine {\it hastrom}.

\begin{figure*}
  \sidecaption
  \includegraphics[width=7cm]{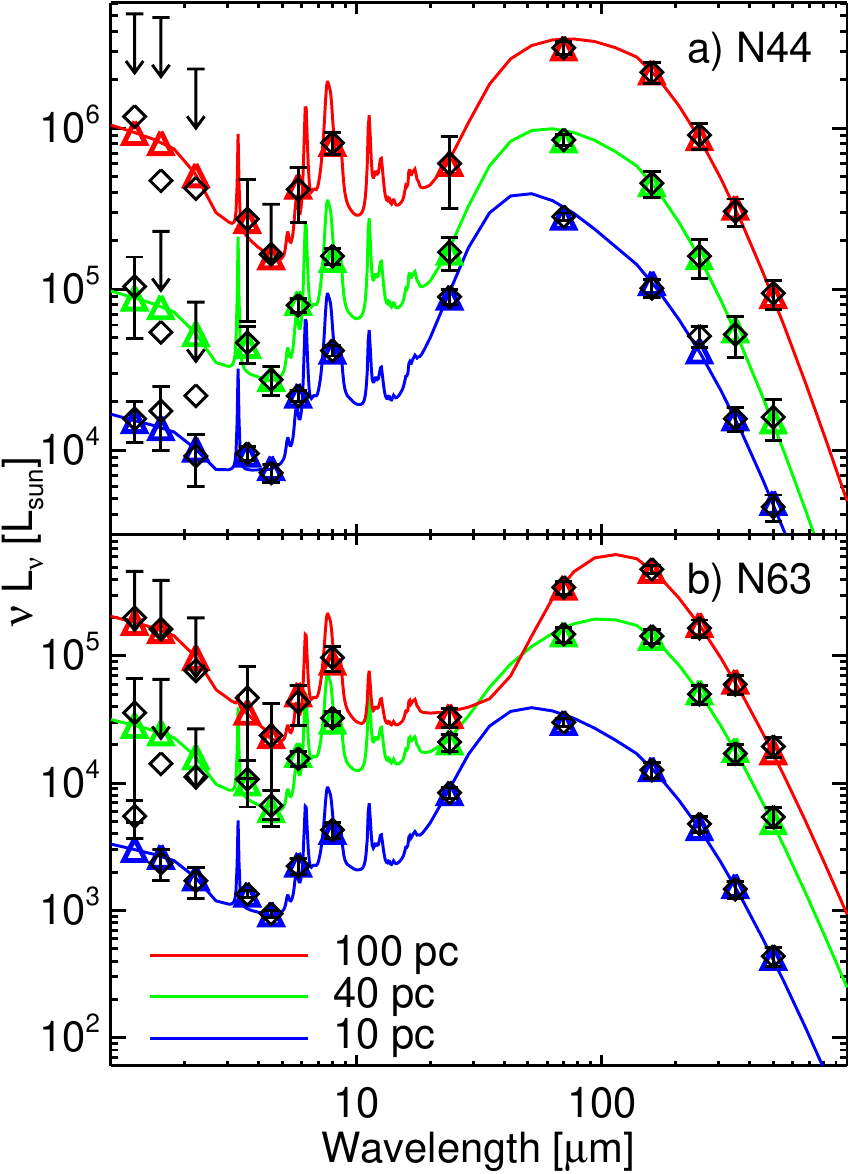}
  \includegraphics[width=7cm]{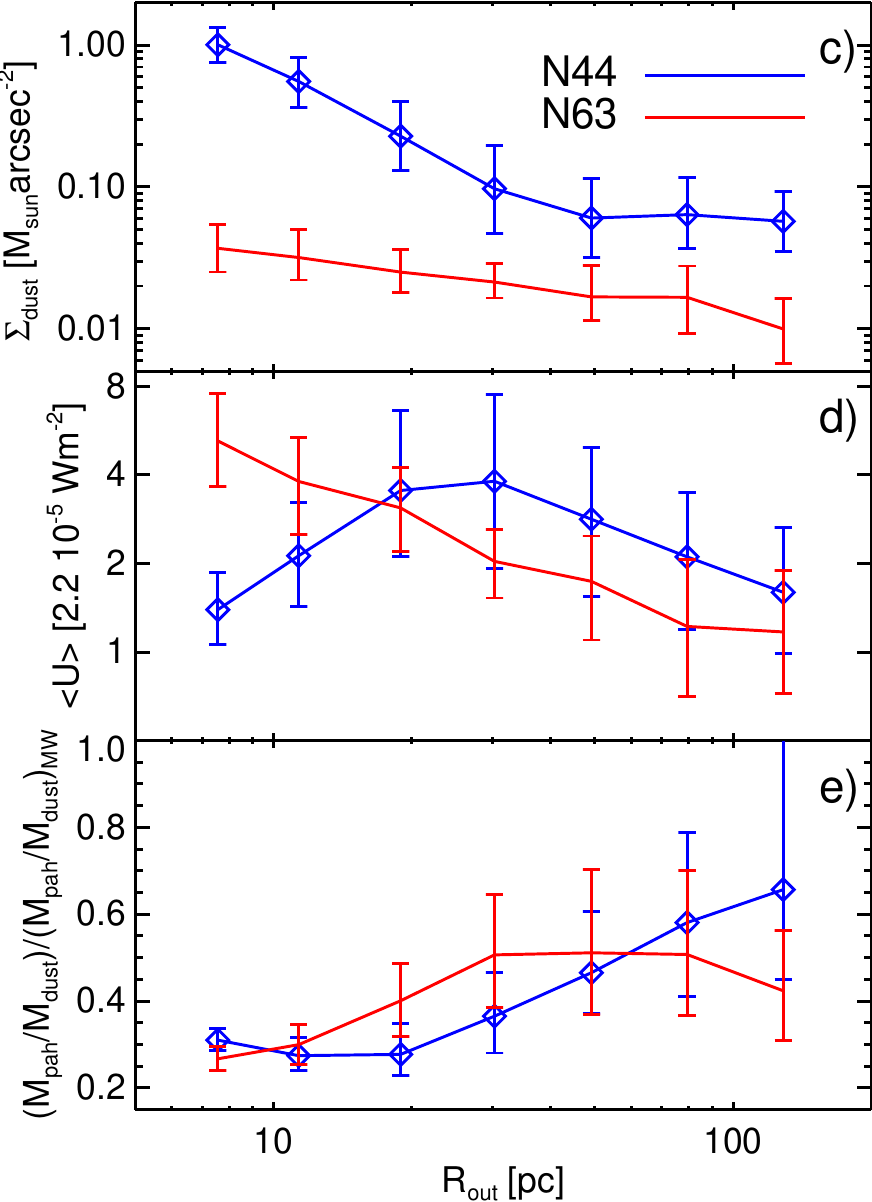}
  \caption{Summary of main results. SEDs of the two SF regions we
    study (left). The black symbols show the integrated photometry of
    three different annuli, lines are the best fit models and the
    coloured triangles represent the synthetic photometry in the
    corresponding filters. The derived dust column density (panel~{\bf
      c}), mean radiation field energy density (panel~{\bf d}), and
    the PAH mass fraction (panel~{\bf e}) as a function of annulus
    size are shown on the right. The error-bars on the parameters in
    Panels {\bf c,d,e} have been derived by propagating the variance
    on the photometry using a Monte-Carlo method (see
    Sec.~\ref{sec:data-treatm-model}). }
  \label{figResults}
\end{figure*}

The center of the HMSFS is determined, by fitting a 2D Gaussian
profile, as the brightest source in the total IR (TIR) image. The TIR
image is obtained by simple integration from 8 to 500~$\mu$m. The
reasoning behind using this definition of the center is that this
location (on a size scale of tens of parsecs ) probably hosts the most
active site of embedded star formation. We extract flux densities (in
Jy) for annuli around the given center (see Fig.~\ref{figMaps}). We
use the following radii for the sizes of the annuli: 35, 47, 78, 125,
200, 330 and 530$^{\prime\prime}$, corresponding to linear sizes of 8,
12, 19, 30, 48, 80 and 128~pc assuming a distance to the LMC of 50~kpc
\citep[e.g.][]{2008AJ....135..112S}. The annuli were chosen to be
larger than the apparent size of the HMSFS so as to be able to also
study the regime in which the emission becomes dominated by the
general LMC. Examples of the extracted SEDs for each region in
different annuli are shown in Fig.~\ref{figResults}a,b. There is a
clear and systematic trend for the far-IR to peak at longer wavelength
for the outer annuli. In the case of N44 this trend is to a large part
offset by the broadness of the far-IR peak. This is an indication that
the emission arises from a broad temperature distribution. We also
note that we detect polycyclic aromatic hydrocarbon (PAH) emission (at
8~$\mu$m) in all SEDs. Since we are, to first order, interested in the
``shape'' of the IR SED as a function of distance, we create relative
flux density maps by dividing the maps at each wavelength by the TIR
map in order to measure the scatter. The scatter inside each annulus
of the relative intensity map was used to estimate the uncertainty
(1$\sigma$) on the flux densities. Note, that this method for
determining the variance causes a large scatter in the near-IR pixels
which are dominated by stellar light. This is the reason for the large
error-bars at these wavelengths in Fig.~\ref{figResults}a,b.

We use a phenomenological dust model, which is described in detail in
\citet{2009A&A...508..645G}, to interpret the observed SEDs. This
model consists of a quantity of dust, with a realistic composition
(PAHs, silicate and graphite) and grain-size distribution (complex
molecules, very small (VSG) and big grains) illuminated by a radiation
field with a range of intensities. The main parameters that we aim to
constrain are: {\it 1)} total dust mass, {\it 2)} the range of
illumination intensities, and {\it 3)} the relative composition, in
particular the mass fraction of PAHs. The radiation field distribution
is represented by a power-law function that describes which fraction
of the matter is exposed to what radiation field
\citep{2002ApJ...576..159D}:
\begin{eqnarray}
  &&M_\mathrm{dust} \propto U^{-\alpha}: U_\mathrm{min} < U < U_\mathrm{max}
  \label{eqn:dale} \\
  &&<{\mathrm U}>  =
  \frac{1}{M_\mathrm{dust}}
  \int_{U_\mathrm{min}}^{U_\mathrm{max}} U
  \frac{{\mathrm d} M_\mathrm{dust}(U)}{{\mathrm d} U} {\mathrm d} U,
\end{eqnarray}
where $U$ is the intensity of the incident radiation field the dust is
subjected to ($U$=1 for the diffuse MW environment), $U_\mathrm{min}$
and $U_\mathrm{max}$ the minimum and maximum value of $U$ and $<$U$>$
is the mass-weighted averaged radiation field intensity.
 
The starting point for the modeling is to adopt dust properties that
fit the MW IR emission well \citep[][distribution
BARE-GR-S]{2004ApJS..152..211Z}. The observed 24~$\mu$m fluxes in the
diffuse ISM force us to use a dust size distribution that is different
from the MW values. The best fits are obtained by reducing the mass
fraction of VSGs to total dust by 50 per cent. We use the MW
interstellar radiation field as the shape of the radiation field and
do not vary this. The derived dust masses are robust against the
choice of interstellar radiation field and the mass fraction of VSGs.
However, the relative mass fractions of the various dust constituents
may depend on these choices.

In comparing the dust mass tracers with the gas tracers
\citep{meixner_special_issue,gordon_special_issue,duval_special_issue}
some issues have been raised about the applicability of this
composition to the LMC, in particular the graphitic component. We have
verified that the derived mass and radiation field profiles (the shape
as a function of annulus) are not sensitive to the choice of the
carbon bearing grains. However, the absolute values of the derived
parameters depend on this choice. Uncertainties on the derived
parameters were estimated using a Monte-Carlo evaluation. The fitting
routine was repeated 300 times with the observational constraints
varied randomly, according to their standard deviations and new best
fit parameters are determined. The ensemble of best fit parameters is
used to calculate the error-bars on each parameter (see
Fig.~\ref{figResults}).

\section{Results}
\label{sec:results}
Fig.~\ref{figResults} {\bf a,b} show representative SEDs and the best
fit models. The models fit the data very well over the entire
wavelength range, for all annuli. Most parameters are well constrained
with a distribution of best fit parameters which is roughly symmetric
around the central value. One exception to this is the $<$U$>$ in the
two outer annuli of N44. We find a distribution around the mean which
is heavily skewed to low values of $<$U$>$. We do not detect a
systematic submm excess, i.e. the whole wavelength range up to
500~$\mu$m is well fit by the standard model. The model and the
500~$\mu$m surface brightness agree on average within 3\% with a very
small dispersion. The model is in accordance with the findings of
\citet{gordon_special_issue}, who show that the evidence in the LMC
for a submm excess is weak and if present it is confined to the more
diffuse and fainter environments.

The SED fits allow us to study the derived properties as a function of
annulus (projected distance). We show in Fig.~\ref{figResults}{\bf
  c,d,e} as a function of radius, the dust column density
($\Sigma_\mathrm{dust}$), average radiation field intensity ($<$U$>$)
and the mass fraction of PAHs normalised to the MW value. Both sources
show a column density profile that decreases outwards, although the
column density in the central region of N63 is not much higher than in
the outer annuli. N63 causes an increase of a factor 2$-$3 of column
density. The column density towards the central regions of N44 is
$\sim$10 times higher than its environment.

The width of the IR SED, which is well determined using the {\it
  SPIRE} data, requires a significant dust mass at lower temperatures.
This is reflected in the low values of $<$U$>$ in panel {\bf d}.
$<$U$>$ exhibits only a small range for the entire sample of SEDs.
$<$U$>$ decreases as a function of distance for N63. N44 does not
exhibit any significant trend and the data are consistent with a
constant $<$U$>$ over a distance scale from 10$-$100~pc. Note that
$<$U$>$ is weighted by dust mass and so it readily traces the
radiation field as seen by the coldest dust along the line of sight.

We calculate the second moment of the radiation field distribution
($\Delta U$), again weighted by dust mass in order to
  quantify the range of radiation field that the matter is exposed to.
  {\it All} SEDs require a wide range of $U$. For N44 $\Delta U$ is
roughly constant at the value of 30 meaning that the entire region is
typified by a radiation field intensity ranging from 1 to 30. N63
shows an outward decreasing $\Delta U$ from $\sim$50 in the inner
annulus to 2 in the outer. Thus the radiation field in N63 spans 5-55
($<$U$>$-$<$U$>$+$\Delta U$) on the inside and 1-3 in the
outer annuli. This implies that the outer annuli in N63 closely
resemble a diffuse environment.

The fraction of mass contained in PAHs increases with increasing
radius out to about 50 per cent of the MW value. We find a significant
depletion of the PAHs towards the central regions over $\sim$20 and
40~pc, for N44 and N63 respectively. Note that the smaller size of the
depleted region is consistent with the fact that N63 is less prominent
compared to its surroundings (see below). Interestingly, the radius of
the depleted environment corresponds well to the area occupied by the
ionising stars in Fig.~\ref{figMaps}. The fraction of the ionised PAHs
is not well constrained but we have checked that there is no
systematic effect of the fitting procedure that causes the observed
trend in Fig.~\ref{figResults}{\bf e}.

\begin{figure}
  \centering
  \includegraphics[width=6cm]{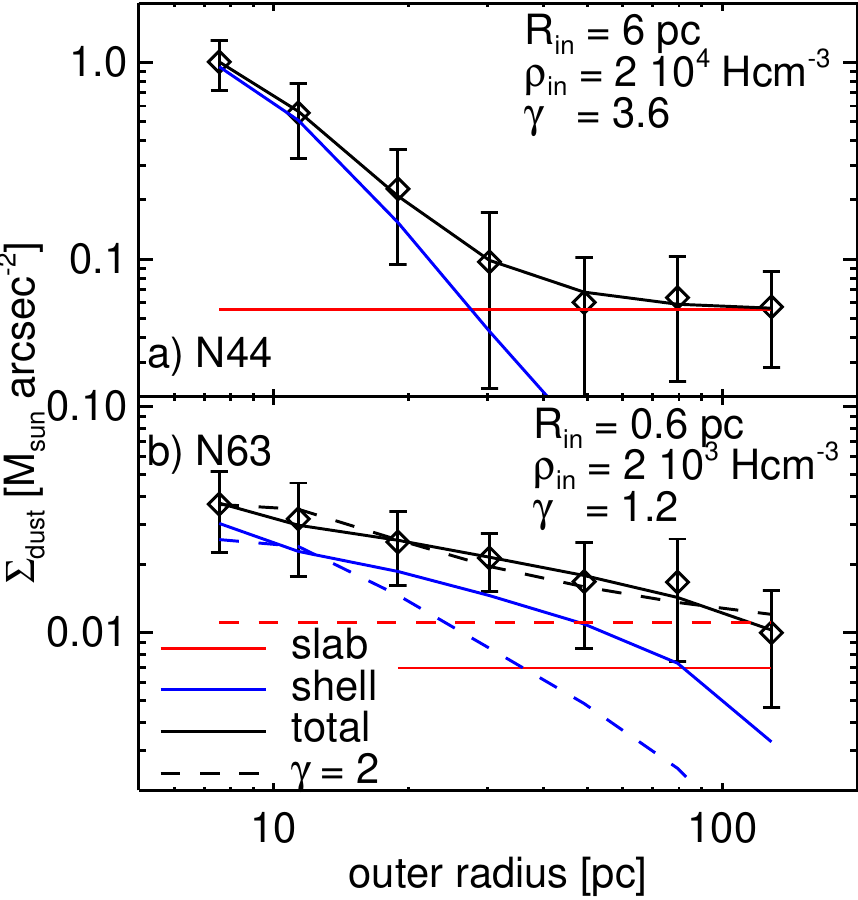}
  \caption{Derived geometrical parameters of the two regions. We show
    the dust column density and the fitted profile of a thick shell,
    with a power-law density profile (blue line) and a slab (red). The
    dashed model corresponds to the case when we fix the exponent of
    the power-law to 2 (see text).}
  \label{geometry}
\end{figure}

We construct a simple geometrical model to try to constrain the size
and density of the matter in the HMSFSs, i.e. to deproject the
observed column densities. We represent the molecular cloud around the
H~{\sc ii}~region as a geometrically thick shell. To this we add a
slab that represents the disk of the LMC, i.e the diffuse extended
emission. We assume a constant dust column density for the slab and a
power-law profile for the shell
($\rho_\mathrm{dust}=\rho_\mathrm{dust,in}(r/R_\mathrm{in})^{-\gamma}$).

The dust mass profile and the best fit decomposition into these two
components and the corresponding parameters are shown in
Fig.~\ref{geometry}. We have propagated the uncertainties on the dust
column density in each annulus (see Fig.~\ref{geometry}). The
uncertainties are significant. We find in case N63, for which the
over-density is weak compared to the slab (i.e. the disk of the LMC),
that the geometrical parameters are not very well constrained. In
particular, it is hard to constrain the $\gamma$ parameter. The best
fit model has $\gamma=1.2$ which seems flat. If we fix $\gamma$ at 2,
the value expected for a cloud in hydrostatic equilibrium, we find
that the distance where the slab component starts to dominate, i.e.
the intersection of the blue and red lines in Fig.~\ref{geometry},
reduces from $\sim$80 to $\sim$30~pc.

\section{Discussion \& Conclusion}
\label{sec:disc--concl}
Figs.~\ref{figResults} and \ref{geometry} show the power of {\it
  Herschel} to determine the matter distribution around HMSFSs. We
investigate the effect the new SPIRE constraints on the derived
parameters and their uncertainties by also fitting the SEDs without
using these SPIRE data. We find that the column density of dust is
often very discrepant (by more than an order of magnitude) from the
values derived using the SPIRE data. The derived $\Sigma_{dust}$
profile for N44 (like Fig.~\ref{figResults} top panel) is virtually
constant at a value of 0.1~M$_{\sun}$\,arcsec$^{-2}$ due to the fact
that the cooler dust, is not well traced by the MIPS data. For N63 the
model over-predicts the derived dust masses for the annuli between 20
to 60 pc by a large factor ($\sim$5). In these cases the model tends
to find very small values of $U_\mathrm{min}$. This artifact could
perhaps be circumvented, when modelling those kind of regions, in
cases where submm constraints are missing, by limiting the allowed
range of $U_\mathrm{min}$ in Eq.~(\ref{eqn:dale}). The difficulty of
constraining the radiation field parameters without the SPIRE bands is
also reflected in the derived uncertainties. In particular the value
of $U_\mathrm{min}$ is ill-constrained which results in large,
asymmetric error-bars on the derived column densities.

N44 and N63 exhibit strikingly different behaviour in the radiation
intensity profile (Fig.~\ref{figResults}{\bf d}.) The lack of a
systematic decrease of $<$U$>$ around N44 indicates that we are
observing dust with a wide range of temperatures along each
line-of-sight. The inner annuli in N44 are affected by the superbubble
to the NE of the OB association, where high values of $<$U$>$ are
expected. The low values of $<$U$>$ for such a luminous SF region may
reflect clumpiness. The profile is clearly incompatible with a
centrally illuminated optically thin irradiation profile. It is clear
from Fig.~\ref{figMaps} that the studied regions are not very
spherically symmetric (azimuthally smooth). In particular, N44
harbours several clusters and the peak of the X-ray emission is
located in a cavity, $\sim$ 20$-$30~pc away from the TIR peak (see
Fig.~\ref{figMaps}). Measuring the azimuthally averaged properties
smears out some of the characteristics. This smearing could have been
the cause for the lack of trend seen in the average $U$ as seen by the
dust (Fig.~\ref{figResults}). We have verified that this small range
of $<$U$>$ is not simply an artifact of this averaging or a wrong
choice of center of the annuli by studying the parameters we derive
pixel by pixel in the maps which makes no assumptions about the
geometry. Indeed the highest $<$U$>$ in N44 is found close to the
center we chose. Except for the very center all other values, with
their scatter, are within the range as depicted in
Fig.~\ref{figResults}. We conclude that the choice of center does not
dominate the lack of trend of N44 in the average radiation field. Thus
this lack of trend reflects the true broad range of irradiation
conditions along all lines of sight in N44 , which is an indication of
the inhomogeneity of the ISM around N44. A simple dust model shows a
deficit in PAHs toward the centers of these two regions. We find no
evidence for a submm excess. We have used the observed dust column
densities surrounding N44 and N63 to derive a 3-D model for these
regions for the first time.

\begin{acknowledgements}
  We acknowledge financial support from the NASA Herschel Science
  Center, JPL contracts \# 1381522 \& 1381650. We thank the
  contributions and support from the European Space Agency (ESA), the
  PACS and SPIRE teams, the Herschel Science Center and the NASA
  Herschel Science Center (esp. A. Barbar and K. Xu) and the PACS and
  SPIRE instrument control centers, without which none of this work
  would be possible. We thank the referee, Glenn White, for comments
  that have improved the paper.
\end{acknowledgements}

\bibliographystyle{aa}
\bibliography{p14628}

\begin{thebibliography}{25}
\expandafter\ifx\csname natexlab\endcsname\relax\def\natexlab#1{#1}\fi

\bibitem[{{Bica} {et~al.}(1996){Bica}, {Claria}, {Dottori}, {Santos}, \&
  {Piatti}}]{1996ApJS..102...57B}
{Bica}, E., {Claria}, J.~J., {Dottori}, H., {Santos}, Jr., J.~F.~C., \&
  {Piatti}, A.~E. 1996, \apjs, 102, 57

\bibitem[{{Calzetti} {et~al.}(2007){Calzetti}, {Kennicutt}, {Engelbracht},
  {Leitherer}, {Draine}, {Kewley}, {Moustakas}, {Sosey}, {Dale}, {Gordon},
  {Helou}, {Hollenbach}, {Armus}, {Bendo}, {Bot}, {Buckalew}, {Jarrett}, {Li},
  {Meyer}, {Murphy}, {Prescott}, {Regan}, {Rieke}, {Roussel}, {Sheth}, {Smith},
  {Thornley}, \& {Walter}}]{2007ApJ...666..870C}
{Calzetti}, D., {Kennicutt}, R.~C., {Engelbracht}, C.~W., {et~al.} 2007, \apj,
  666, 870

\bibitem[{{Caulet} {et~al.}(2008){Caulet}, {Gruendl}, \&
  {Chu}}]{2008ApJ...678..200C}
{Caulet}, A., {Gruendl}, R.~A., \& {Chu}, Y. 2008, \apj, 678, 200

\bibitem[{{Chen} {et~al.}(2009){Chen}, {Chu}, {Gruendl}, {Gordon}, \&
  {Heitsch}}]{2009ApJ...695..511C}
{Chen}, C., {Chu}, Y., {Gruendl}, R.~A., {Gordon}, K.~D., \& {Heitsch}, F.
  2009, \apj, 695, 511

\bibitem[{{Dale} \& {Helou}(2002)}]{2002ApJ...576..159D}
{Dale}, D.~A. \& {Helou}, G. 2002, \apj, 576, 159

\bibitem[{Duval {et~al.}(2010)}]{duval_special_issue}
Duval, J. {et~al.} 2010, \aap, this volume

\bibitem[{{Fazio} {et~al.}(2004){Fazio}, {Hora}, {Allen}, {Ashby}, {Barmby},
  {Deutsch}, {Huang}, {Kleiner}, {Marengo}, {Megeath}, {Melnick}, {Pahre},
  {Patten}, {Polizotti}, {Smith}, {Taylor}, {Wang}, {Willner}, {Hoffmann},
  {Pipher}, {Forrest}, {McMurty}, {McCreight}, {McKelvey}, {McMurray}, {Koch},
  {Moseley}, {Arendt}, {Mentzell}, {Marx}, {Losch}, {Mayman}, {Eichhorn},
  {Krebs}, {Jhabvala}, {Gezari}, {Fixsen}, {Flores}, {Shakoorzadeh}, {Jungo},
  {Hakun}, {Workman}, {Karpati}, {Kichak}, {Whitley}, {Mann}, {Tollestrup},
  {Eisenhardt}, {Stern}, {Gorjian}, {Bhattacharya}, {Carey}, {Nelson},
  {Glaccum}, {Lacy}, {Lowrance}, {Laine}, {Reach}, {Stauffer}, {Surace},
  {Wilson}, {Wright}, {Hoffman}, {Domingo}, \& {Cohen}}]{2004ApJS..154...10F}
{Fazio}, G.~G., {Hora}, J.~L., {Allen}, L.~E., {et~al.} 2004, \apjs, 154, 10

\bibitem[{{Galametz} {et~al.}(2009){Galametz}, {Madden}, {Galliano}, {Hony},
  {Schuller}, {Beelen}, {Bendo}, {Sauvage}, {Lundgren}, \&
  {Billot}}]{2009A&A...508..645G}
{Galametz}, M., {Madden}, S., {Galliano}, F., {et~al.} 2009, \aap, 508, 645

\bibitem[{Gordon {et~al.}(2010)}]{gordon_special_issue}
Gordon, K. {et~al.} 2010, \aap, this volume

\bibitem[{{Gordon} {et~al.}(2008){Gordon}, {Engelbracht}, {Rieke}, {Misselt},
  {Smith}, \& {Kennicutt}}]{2008ApJ...682..336G}
{Gordon}, K.~D., {Engelbracht}, C.~W., {Rieke}, G.~H., {et~al.} 2008, \apj,
  682, 336

\bibitem[{Griffin {et~al.}(2010)}]{griffin_special_issue}
Griffin, M. {et~al.} 2010, \aap, this volume

\bibitem[{{Henize}(1956)}]{1956ApJS....2..315H}
{Henize}, K.~G. 1956, \apjs, 2, 315

\bibitem[{{Kennicutt}(1998)}]{1998ARA&A..36..189K}
{Kennicutt}, Jr., R.~C. 1998, \araa, 36, 189

\bibitem[{{Lee} {et~al.}(2009){Lee}, {Gil de Paz}, {Tremonti}, {Kennicutt},
  {Salim}, {Bothwell}, {Calzetti}, {Dalcanton}, {Dale}, {Engelbracht}, {Funes},
  {Johnson}, {Sakai}, {Skillman}, {van Zee}, {Walter}, \&
  {Weisz}}]{2009ApJ...706..599L}
{Lee}, J.~C., {Gil de Paz}, A., {Tremonti}, C., {et~al.} 2009, \apj, 706, 599

\bibitem[{{Lucke} \& {Hodge}(1970)}]{1970AJ.....75..171L}
{Lucke}, P.~B. \& {Hodge}, P.~W. 1970, \aj, 75, 171

\bibitem[{{Madden} {et~al.}(2006){Madden}, {Galliano}, {Jones}, \&
  {Sauvage}}]{2006A&A...446..877M}
{Madden}, S.~C., {Galliano}, F., {Jones}, A.~P., \& {Sauvage}, M. 2006, \aap,
  446, 877

\bibitem[{Meixner {et~al.}(2010)}]{meixner_special_issue}
Meixner, M. {et~al.} 2010, \aap, this volume

\bibitem[{{Oey}(1996)}]{1996ApJS..104...71O}
{Oey}, M.~S. 1996, \apjs, 104, 71

\bibitem[{{Oey} \& {Massey}(1995)}]{1995ApJ...452..210O}
{Oey}, M.~S. \& {Massey}, P. 1995, \apj, 452, 210

\bibitem[{Pilbratt {et~al.}(2010)}]{pilbratt_special_issue}
Pilbratt, G. {et~al.} 2010, \aap, this volume

\bibitem[{{Rieke} {et~al.}(2004){Rieke}, {Young}, {Engelbracht}, {Kelly},
  {Low}, {Haller}, {Beeman}, {Gordon}, {Stansberry}, {Misselt}, {Cadien},
  {Morrison}, {Rivlis}, {Latter}, {Noriega-Crespo}, {Padgett}, {Stapelfeldt},
  {Hines}, {Egami}, {Muzerolle}, {Alonso-Herrero}, {Blaylock}, {Dole}, {Hinz},
  {Le Floc'h}, {Papovich}, {P{\'e}rez-Gonz{\'a}lez}, {Smith}, {Su}, {Bennett},
  {Frayer}, {Henderson}, {Lu}, {Masci}, {Pesenson}, {Rebull}, {Rho}, {Keene},
  {Stolovy}, {Wachter}, {Wheaton}, {Werner}, \&
  {Richards}}]{2004ApJS..154...25R}
{Rieke}, G.~H., {Young}, E.~T., {Engelbracht}, C.~W., {et~al.} 2004, \apjs,
  154, 25

\bibitem[{{Schaefer}(2008)}]{2008AJ....135..112S}
{Schaefer}, B.~E. 2008, \aj, 135, 112

\bibitem[{{Skrutskie} {et~al.}(2006){Skrutskie}, {Cutri}, {Stiening},
  {Weinberg}, {Schneider}, {Carpenter}, {Beichman}, {Capps}, {Chester},
  {Elias}, {Huchra}, {Liebert}, {Lonsdale}, {Monet}, {Price}, {Seitzer},
  {Jarrett}, {Kirkpatrick}, {Gizis}, {Howard}, {Evans}, {Fowler}, {Fullmer},
  {Hurt}, {Light}, {Kopan}, {Marsh}, {McCallon}, {Tam}, {Van Dyk}, \&
  {Wheelock}}]{2006AJ....131.1163S}
{Skrutskie}, M.~F., {Cutri}, R.~M., {Stiening}, R., {et~al.} 2006, \aj, 131,
  1163

\bibitem[{{Soifer} {et~al.}(1986){Soifer}, {Sanders}, {Neugebauer},
  {Danielson}, {Lonsdale}, {Madore}, \& {Persson}}]{1986ApJ...303L..41S}
{Soifer}, B.~T., {Sanders}, D.~B., {Neugebauer}, G., {et~al.} 1986, \apjl, 303,
  L41

\bibitem[{{Zubko} {et~al.}(2004){Zubko}, {Dwek}, \&
  {Arendt}}]{2004ApJS..152..211Z}
{Zubko}, V., {Dwek}, E., \& {Arendt}, R.~G. 2004, \apjs, 152, 211

\end{thebibliography}

\end{document}